# Ultra-sharp magnetization steps in perovskite manganites

R. Mahendiran,[1] A. Maignan,[2] S. Hébert,[2] C. Martin,[2] M. Hervieu,[2] B. Raveau,[2] J. F. Mitchell,[3] and P. Schiffer[1]

[1]*Department of Physics and Materials Research Institute, Pennsylvania State University, University Park, PA 16802*

[2]*Laboratoire CRISMAT, ISMRA, 6 Boulevard du Maréchal Juin, 14050 Caen Cedex, France*

[3]*Materials Science Division, Argonne National Laboratory, Argonne, IL 60439*

**Abstract**

We report a detailed study of step-like metamagnetic transitions in the magnetization and resistivity of polycrystalline $Pr_{0.5}Ca_{0.5}Mn_{0.95}Co_{0.05}O_3$. The steps have a sudden onset below a critical temperature, are extremely sharp (width $< 2 \times 10^{-4}$ T), and occur at critical fields which are linearly dependent on the absolute value of the cooling field in which the sample is prepared. Similar transitions are also observed at low temperature in non-Co doped manganites, including single crystal samples. These data show that the steps are an intrinsic property, qualitatively different from either previously observed higher temperature metamagnetic transitions in the manganites or metamagnetic transitions observed in other materials.

Metamagnetic phase transitions, those induced by a magnetic field, have an extended history of investigation in long range ordered antiferromagnets [1]. Much recent research has also focused on a wide range of other material systems including clean metals [2], magnetic cluster compounds [3], and geometrically frustrated magnets [4], all of which display transitions induced by a magnetic field. Included in this category are many of the perovskite "colossal magnetoresistance" manganites [5] which display first order field-induced transitions in an applied field at low temperatures from an antiferromagnetic charge-orbital ordered insulating (COOI) phase to a ferromagnetic metallic phase. In the case of the manganites, the phases typically coexist in low magnetic fields [6] (particularly in the presence of Mn site doping [7,8,9,10,11]), and the metamagnetic transitions progress through an increasing fraction of the ferromagnetic phase in the phase-separated materials [12].

Metamagnetism in the manganites has been investigated primarily at temperatures above 5 K, and the transitions are rather broad in that temperature regime as may be expected from the inhomogeneous nature of the low field phase-separated state. There has, however, been a recent report of unusual step-like magnetic field dependence of the magnetization and other properties measured at temperatures below 5 K in a particular series of compounds, $Pr_{0.5}Ca_{0.5}MnO_3$ doped with a few percent of other cations such as Sc or Ga on the Mn site [13]. We present here a detailed investigation of the temperature and field dependence of reproducible steps in the magnetization and resistivity of $Pr_{0.5}Ca_{0.5}Mn_{1-x}Co_xO_3$ ($x = 0.05$). We find that the onset of the steps is extremely sharp in temperature and that the steps themselves are extraordinarily sharp with a width of $\Delta H/H \sim 10^{-4}$ even at temperatures immediately below the onset temperature. Surprisingly, the field ($H_c$) at which the steps occur increases after field cooling, and $H_c$ itself is linear in the magnitude of the cooling field. We also observe a similar step transition in manganites which are not doped on the Mn site, demonstrating that the low temperature steps comprise a qualitatively new class of metamagnetic transitions in the manganites which also differs significantly from those seen in other magnetic systems.

We have studied high quality polycrystalline samples of $Pr_{0.5}Ca_{0.5}Mn_{1-x}Co_xO_3$ prepared by a standard solid state technique (sample preparation and characterization details are published elsewhere [7]). We measured magnetization ($M$) on a Quantum Design SQUID magnetometer and resistivity ($\rho$) by a standard four-probe technique. Data are shown primarily for $x = 0.05$ (chosen because it is unaffected by thermal cycling unlike previously studied compounds exhibiting steps [13] or Co-doped materials with $x < 0.04$), but the results are qualitatively equivalent for $x = 0.06$ and $0.07$ [14]. The results have been qualitatively reproduced in three separate series of samples grown under slightly different conditions and in two different sample preparation laboratories.

In figure 1 we show the high-temperature behavior of this material through the temperature dependence of the magnetization, $M(T)$, measured on heating after the sample was either zero-field cooled (ZFC) or field cooled (FC). As seen in the inset to figure 1, there is a kink in $H/M(T)$ around 220 K, which is a signature of the onset of charge-orbital ordering [10]. Both the ZFC and FC $M(T)$ show a ferromagnetic transition at $T_C \sim 60$ K in low fields. We presume, as in the case of many other manganites, that charge ordering coexists with ferromagnetism in the low temperature phase. The coexistence of the two phases in an inhomogeneous state at low temperatures is evidenced [12] by a divergence of the ZFC $M(T)$ at $H = 1$ T from the FC curve below $T_{irr}$ ~ 30 K ($T_{irr}$ is not suppressed to zero even at $H = 7$ T, implying that this is not a homogeneous spin glass phase).

The two-phase nature of the low temperature phase is also evident in figure 2a where we plot $M(H)$ at $T = 5$ K. There is an initial increase in $M(H)$, attributable to alignment of the ferromagnetic regions of the sample, and then a kink at $H \sim 2$ T followed by a second increase at higher fields associated with conversion of the charge-ordered regions to a ferromagnetic conducting phase (similar behavior has been observed in other mixed phase manganites such as $Pr_{0.5}Ca_{0.5}Mn_{0.98}Cr_{0.02}O_3$ [11] and $La_{0.250}Pr_{0.375}Ca_{0.375}MnO_3$ [15]). The magnetization appears to saturate at 7 T, with a saturation moment of 3.3 $\mu_B$/formula unit -- close to the 3.4 $\mu_B$/formula unit expected for $Pr_{0.5}Ca_{0.5}Mn^{3+}_{0.40}Co^{2+}_{0.05}Mn^{4+}_{0.55}O_3$. When the field is reduced from 7 T, $M(H)$ behaves like a long range ferromagnet with little change down to 2 T and then a rapid decrease as $H \rightarrow 0$.

The breadth of the 5 K field-induced transition from a predominantly COOI to predominantly ferromagnetic metallic state suggests a wide distribution of critical fields which drive this transition in different parts of the sample. This is consistent with the presumption of phase separation and the behavior in other manganite materials in which such a transition has been observed [11,12,15,16]. The behavior changes dramatically at lower temperatures, however, as shown in figures 2b-d. There is no qualitative change in $M(H)$ down to 4.7 K, as shown in figure 2b, but at 4.6 K $M(H)$ shows an abrupt step near $H_{c1} = 2.5$ T. The relative magnitude of the step can be written as $g = \Delta M/M_{sat} = 0.18$ at 4.6 K. After the step, $M(H)$ increases linearly with a relatively small slope until it merges with the $M(H)$ curve recorded at $T = 4.7$ K. Such steps are also observed at lower temperatures, and additional steps appear with decreasing temperature. The ZFC $M(H)$ at $T = 4$ K (figure 2c) again shows a single step at $H_{c1} = 2.4$ T of higher magnitude ($g = 0.43$). At $T = 3$ K, however, there are two sharp step transitions in the ZFC data, at $H_{c1} = 2.3$ T ($g = 0.41$) and at $H_{c2} = 4.5$ T ($g = 0.22$), as seen in figure 2d. At all temperatures below 5 K, the saturation magnetization at $H = 7$ T is close to the theoretical limit and the magnetization behaves like a homogeneous ferromagnet upon reducing the field from 7 T and on subsequent field sweeps.

The measurements in figure 2a-d were performed with a field interval of 0.2 T, and therefore do not set a strict limit on the width of the steps. We also measured with much smaller field intervals of 0.2 mT (2 Oe) for $T = 2$ and 3 K. As shown in the inset to figure 2d, the step width is even narrower than this interval, indicating that the transition across the entire sample happens at essentially the same magnetic field. These extraordinarily sharp transitions are quite surprising given the polycrystalline nature of the sample and the presumed inhomogeneous nature of the phase-separated low temperature state in these materials.

Since field-cooling changes the relative fraction of ferromagnetic metallic and COOI phases, we also studied samples which were field-cooled with H > 0 T from $T = 120$ K ($> T_C$). After stabilization at the measurement temperature, the field was reduced to zero, and then $M(H)$ was measured up to 7 T and back down to zero field. At $T = 5$ K, such field cooling apparently increases the ferromagnetic fraction of the sample at low fields and thus results in a larger low field magnetization. This behavior is consistent with conventional understanding of phase separation in the manganites, and similar behavior has been seen in $Nd_{0.5}Ca_{0.5}Mn_{0.98}Cr_{0.02}O_3$ [8]. Field cooling similarly increases the low field magnetization at temperatures below the onset of the step transitions, but *the step transitions are shifted to higher fields as a result of field-cooling*, which is surprising since field cooling should enhance ferromagnetic tendencies. The shift of $H_c$ is strikingly linear in the cooling field (as shown in the inset to figure 4b from resistivity data discussed below), but field-cooling in a sufficiently high magnetic field converts the sample into an almost fully ferromagnetic state, eliminating the steps. The steps after field cooling are not measurably broadened beyond our limit of 0.2 mT, implying that the magnetic phase of the sample is uniformly affected by the cooling field. Furthermore, as shown in figure 3, cooling in a negative field and then raising the field continuously to its maximum positive value results in exactly the same shift of $H_c$.

In figure 4a we plot the field dependence of the FC and ZFC resistivity, $\rho(H)$, in analogy to the magnetization data in figure 2d. Note that $\rho(H)$ shows downward steps at the same fields as the $M(H)$ data show upward steps. These data, along with the sizable magnitude of the magnetization steps, indicate that the entire sample is involved in the step transitions, rather than isolated regions. In figure 4b we plot ZFC $\rho(H)$ at several temperatures, and we see that a third step appears in the $T = 2$ K data at $H_{c3} \sim 7.36$ T. Surprisingly, although the steps appear suddenly below a certain temperature, there is very little temperature dependence to $H_{c1}$ or $H_{c2}$. This is clearly seen in the inset to figure 3b where we demonstrate the linear dependence of $H_{c1}$ and $H_{c2}$ on the cooling field, since the data for all of the different measurement temperatures lie on top of each other. A careful measurement of the temperature dependence of $H_{c1}$ indicates that it does decrease slightly (by ~ 0.2 T) from 4.5 K to 2 K [14].

Our data show that the step-like transitions are a qualitatively new phenomenon in the manganites. While previously observed field-induced transitions at higher temperatures are rather broad (width ~ 1 T) and sharpen somewhat with decreasing temperature [16], the steps we observe are extremely sharp, even at the highest temperature at which they are observed. These step-like transitions are not unique to $Pr_{0.5}Ca_{0.5}MnO_3$ doped on the Mn site, but appear to be a generic feature of a broad range of manganites. We also observe the onset of a similar step-like transition in single crystal $Pr_{0.7}Ca_{0.3}MnO_3$ at 2 K (figure 4 inset) and also in both single crystal and polycrystalline $Pr_{0.65}(Ca_{1-y}Sr_y)_{0.35}MnO_3$ and polycrystalline $Sm_{0.5}Sr_{0.5}MnO_3$ although the same cooling field dependence of $H_c$ is not observed in all these materials and the observation of the step at 2 K in $Pr_{0.7}Ca_{0.3}MnO_3$ is also depended on the field sweep rate. The charge-orbital ordered antiferromagnetic domains in the Co doped $Pr_{0.5}Ca_{0.5}MnO_3$ are smaller than in the undoped compounds [7-11], suggesting a possible explanation for why we can observe step transitions at lower fields and higher temperatures in these materials than in the undoped compounds.

There are examples of step-like metamagnetic transitions in other magnetic materials, but those transitions are either not as sharp (i.e. their width depends strongly on temperature), or they occur only in single crystal samples when the applied field is along a particular direction [1,17,18]. Neither case provides a good model for the present behavior, where the steps are extraordinarily sharp even just below the onset temperature and the samples are polycrystalline. One possible explanation for the step transitions in the manganites is that they are due to martensitic effects associated with strain between the phase-separated regions [19], but such a mechanism would have difficulty accounting for the presence of multiple steps which are reproduced in different samples (given the structural and magnetoelectronic inhomogeneity intrinsic to polycrystalline manganites). The sharpness of the steps suggests that they reflect an intrinsic thermodynamic property of the charge-ordered phase, i.e. phase transitions through which the charge-ordered regions are becoming successively more polarized in steps between different canted antiferromagnetic [1] or ferrimagnetic (as in $FeCl_2 \cdot 2H_2O$ [18]) arrangements. Orbital orbiting in the charge-ordered domains may result in such behavior by stabilizing certain types of spin structures with increasing field [20]. Alternatively, it is possible that the dipole interactions between Mn spins provide the anisotropy typically associated with such transitions. One may be able to explain the shift of the steps with cooling field by postulating that field cooling alters the orbital configurations in the COOI domains, making them more difficult to transform into a highly polarized state [20]. The field cooling effect may also be explained as a softening of the antiferromagnetic exchange interaction, which would increase the polarization of the charge-ordered regions and reduce the Zeeman energy advantage of the ferromagnetic state to which the system is transforming.

The important question raised by our data is how the steps can be so sharp and have such a sudden onset temperature in polycrystalline samples with an inhomogeneous phase-separated ground state, since such sharp steps in other metamagnetic systems require a high degree of crystalline anisotropy [1,17]. The reproducibility of the observed phenomena in a range of compounds make it clear that these step transitions in the manganites comprise a robust and qualitatively new sort of metamagnetism, presenting a challenge to the theoretical community working on these materials.

**Acknowledgements:** The authors gratefully acknowledge the support of NSF grant DMR-0101318 and MENRT (France). The submitted manuscript has been created in part by the University of Chicago as Operator of Argonne National Laboratory ("Argonne") under Contract No. W-31-109-ENG-38 with the U.S. Department of Energy. The U.S. Government retains for itself, and others acting on its behalf, a paid-up, nonexclusive, irrevocable worldwide license in said article to reproduce, prepare derivative works, distribute copies to the public, and perform publicly and display publicly, by or on behalf of the Government.

**Figure captions:**

**Figure 1.** Temperature dependence of the magnetization of $Pr_{0.5}Ca_{0.5}Mn_{0.95}Co_{0.05}O_3$ while warming from 2 K after zero field cooling and field cooling. The inset shows $H/M(T)$ in the high temperature range at $H = 1$ T, where a kink indicates $T_{CO}$, the charge-orbital ordering temperature.

**Figure 2.** Field dependence of the magnetization at different temperatures measured while stepping from $H = 0$ T $\rightarrow H = 7$ T $\rightarrow H = 0$ T in 0.2 T steps. **(a).** $T = 5$ K, **(b).** $T = 4.7$ K & 4.6 K, **(c).** $T = 4$ K, and **(d).** $T = 3$ K. Note that the smooth metamagnetic transition at $T = 4.7$ K changes to include an abrupt jump at $T = 4.6$ K. Data are taken either after zero-field-cooling (ZFC) or after field cooling (FC) from 120 K ($> T_C$) in the indicated magnetic fields (the FC data are taken on increasing field, after the field is first reduced to zero at the measurement temperature). The inset in figure 2(d) shows data taken with field interval of 0.2 mT, placing an upper limit on the width of the steps.

**Figure 3.** Comparison of $M(H)$ at $T = 2$K after cooling in positive and negative fields. Positive field-cooled data were taken in the same manner as the data in figure 2.

**Figure 4. (a).** Field dependence of the resistivity at T = 3 K in ZFC and FC modes showing steps which correspond to those observed in the magnetization (figure 2d). The inset shows the onset of a similar step transition in single crystal $Pr_{0.7}Ca_{0.3}MnO_3$ at T = 2 K (sample preparation details given in [12]). **(b).** Field dependence of the resistivity for 2 K $\leq$ T $\leq$ 5 K. The inset shows the cooling field dependence of $H_c$ at different temperatures where the color of the symbols corresponds to the different temperatures in the main figure.

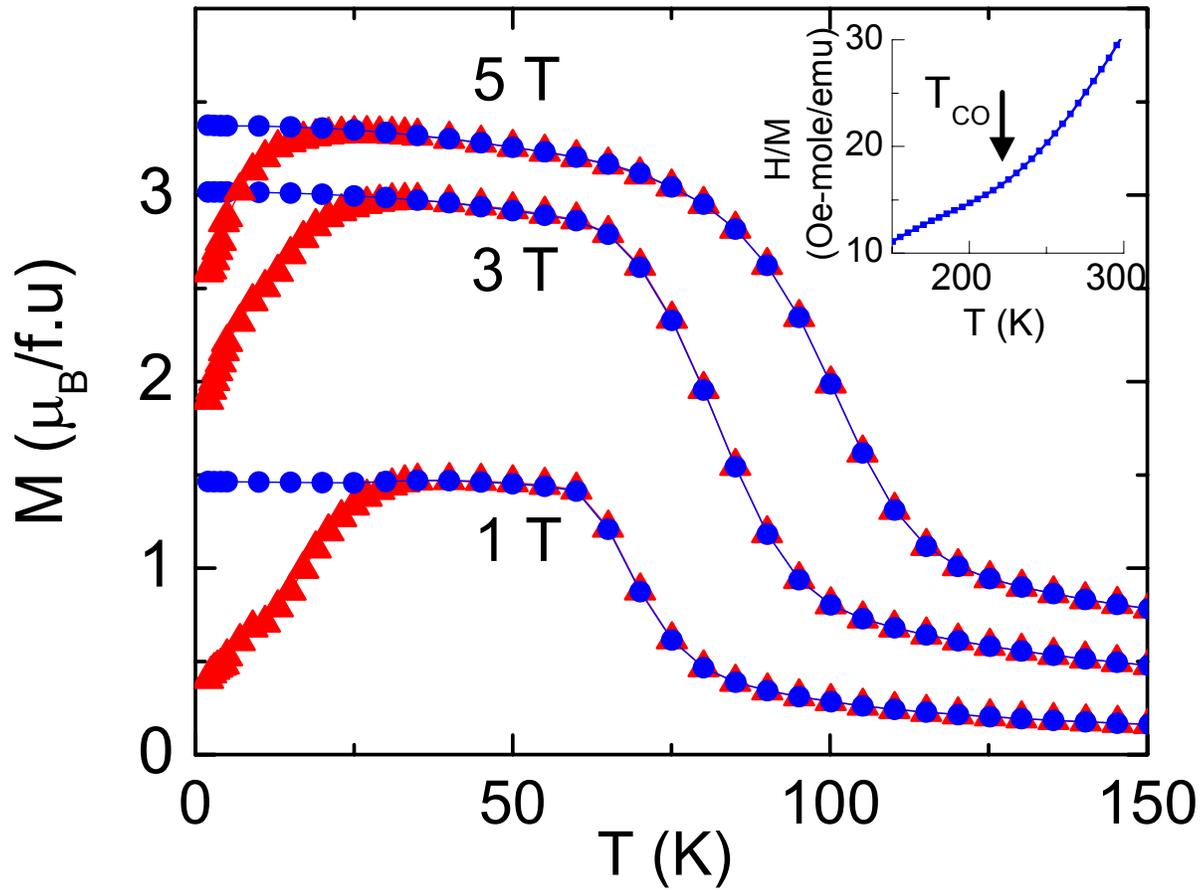

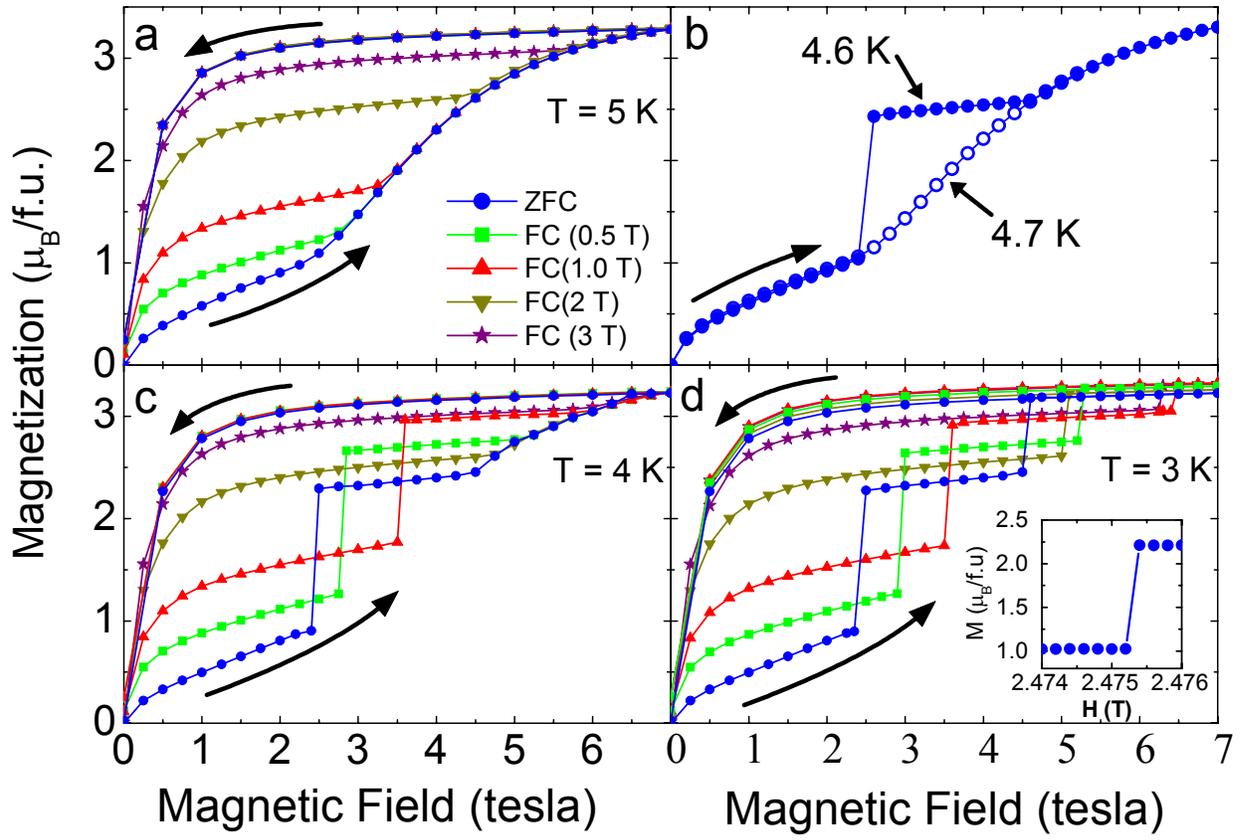

Figure 2 (Mahendiran *et al.*)

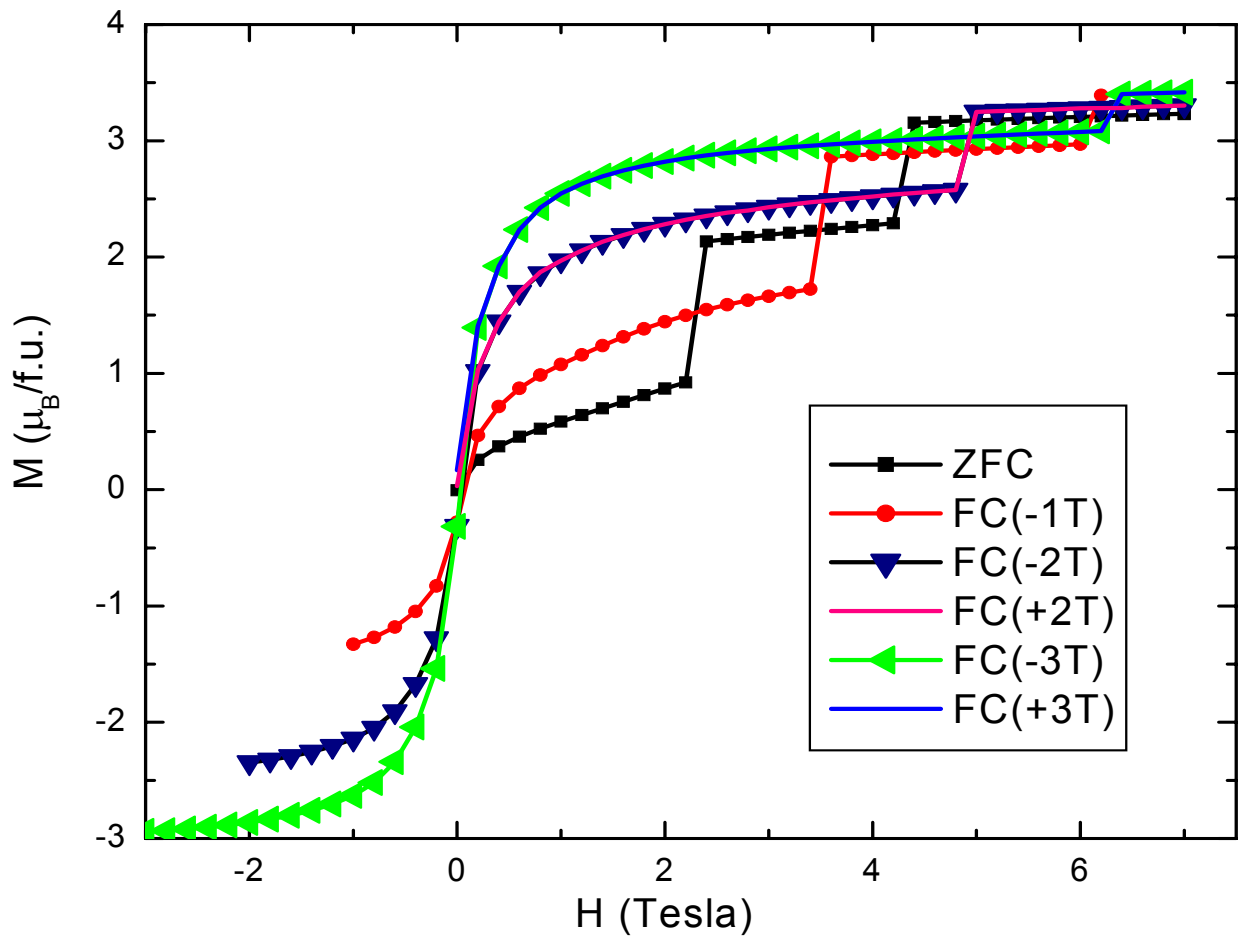

Figure 3 (Mahendiran et al.)

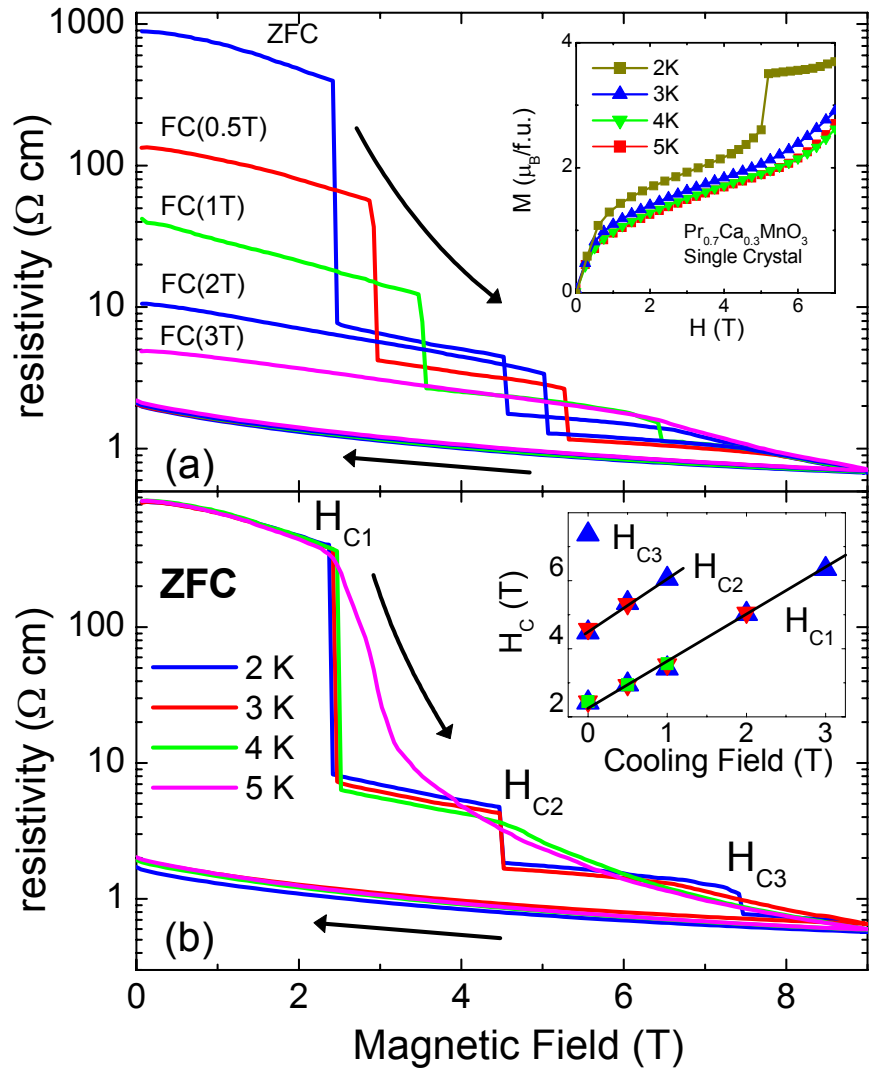